\documentclass[fleqn,usenatbib]{mnras}

\usepackage{newtxtext,newtxmath}

\usepackage[T1]{fontenc}
\usepackage{ae,aecompl}

\usepackage{graphicx}    
\usepackage{amsmath}    
\usepackage{amssymb}    

\title[Comparing the alkali lines on HD189733b]{Probing the atmosphere of HD189733b with the \ion{Na}{i} and \ion{K}{i} lines}
\author[E. Keles et al.]{
E. Keles,$^{1}$
D. Kitzmann,$^{2}$
M. Mallonn,$^{1}$
X. Alexoudi,$^{1}$
L. Fossati,$^{3}$
L. Pino,$^{4,5}$
J. V. Seidel ,$^{6}$
\newauthor
T. A. Carroll,$^{1}$
M. Steffen,$^{1}$
I. Ilyin,$^{1}$
K. Poppenh\"ager,$^{1}$
K. G. Strassmeier$^{1}$
C. von Essen,$^{7}$
\newauthor
V. Nascimbeni,$^{8,9}$
J. D. Turner,$^{10,11}$ \\ \\
$^{1}$Leibniz-Institut f\"ur Astrophysik Potsdam (AIP), An der Sternwarte 16, 14482 Potsdam, Germany\\
$^{2}$University of Bern, Center for Space and Habitability, Gesellschaftsstrasse 6, CH-3012, Bern, Switzerland\\
$^{3}$Space Research Institute, Austrian Academy of Sciences, Schmiedlstr. 6, 8042 Graz, Austria \\
$^{4}$Anton Pannekoek Institute for Astronomy, University of Amsterdam, Science Park 904, 1098 XH Amsterdam, The Netherlands\\
$^{5}$INAF-Osservatorio Astrofisico di Arcetri, Largo Enrico Fermi 5, I-50125 Firenze, Italy\\
$^{6}$Observatoire de l'Universit{\'e} de Gen{\`e}ve, 51 chemin des Maillettes, 1290 Versoix, Switzerland\\
$^{7}$Stellar Astrophysics Centre, Department of Physics and Astronomy, Aarhus University, Ny Munkegade 120, 8000 Aarhus C, Denmark\\
$^{8}$Istituto Nazionale di Astrofisica, Osservatorio Astronomico di Padova, 35122 Padova, Italy\\
$^{9}$Dipartimento di Fisica e Astronomia - Universta di Padova, Vicolo dell Osservatorio 3, I-35122 Padova\\
$^{10}$Cornell University, Ithaca, New York, USA; \\
$^{11}$University of Virginia, Charlottesville, Virginia, USA\\
}

\date{Accepted XXX. Received YYY; in original form ZZZ}

\pubyear{2020}

\begin{document}
\label{firstpage}
\pagerange{\pageref{firstpage}--\pageref{lastpage}}
\maketitle

\begin{abstract}
High spectral resolution transmission spectroscopy is a powerful tool to characterize exoplanet atmospheres. Especially for hot Jupiters, this technique is highly relevant, due to their high altitude absorption e.g. from resonant sodium (\ion{Na}{i}) and potassium (\ion{K}{i}) lines. We resolve the atmospheric \ion{K}{i}-absorption on HD189733b with the aim to compare the resolved \ion{K}{i}-line and previously obtained high resolution \ion{Na}{i}-D-line observations with synthetic transmission spectra. The line profiles suggest atmospheric processes leading to a line broadening of the order of $\sim$10 km/s for the \ion{Na}{i}-D-lines, and only a few km/s for the \ion{K}{i}-line. The investigation hints that either the atmosphere of HD189733b lacks a significant amount of \ion{K}{i} or the alkali lines probe different atmospheric regions with different temperature, which could explain the differences we see in the resolved absorption lines.
\end{abstract}




\section{Introduction}
The characterization of exoplanets and especially their atmospheres increases our understanding of planetary formation and evolution. One possibility to characterize exoplanets is transmission spectroscopy, thus inferring the fingerprints of planetary atmospheres \citep[]{SeagerSasslow2000}, as a part of the starlight is absorbed or scattered by atmospheric constituents during transit. 

Using high-resolution spectroscopy \citep[]{Snellen2010, Cauley2019, Yan2019} one can reveal information about sharp line absorption at high altitudes. These absorptions are investigated usually using either the so called "excess absorption method", where the flux of the spectral range of interest is integrated within a bandwidth and divided by the flux within a reference band \citep{Charbonneau2002} or the "division method", where the observations during the transit are divided by the mean of the out- of- transit observations  \citep{Wyttenbach2015}, showing the spectrally resolved planetary absorption. The information inferred from the spectral lines can be used to determine different properties from the probed atmospheric region e.g. the temperature profile \citep{Wyttenbach2015}, number densities \citep{Huang2017}, escape \citep{Fossati2018}, aerosols \citep{Pino2018,Pino2018b} or the pressure level \citep{Gibson2020,Turner2020}. Absorption lines of interest are the alkali lines \ion{Na}{i} and the \ion{K}{i} with their strong absorption cross-sections \citep{Lavvas2014}, where the exoplanet HD189733b is one of the first exoplanets where atmospheric Na absorption was found \citep{Redfield2008}.
\newline
The exoplanet HD189733b has an apparent radius of \mbox{1.151 $\times$ R\textsubscript{Jup}} and an orbital period of \mbox{~2.21 days}. It orbits an active K-dwarf star with an apparent magnitude of \mbox{V = 7.7 mag} and a radius of \mbox{0.752 $\times$ R\textsubscript{Sun}}. The first \ion{Na}{i} detection in the atmosphere of HD189733b was presented by \citet{Redfield2008} using the High Resolution Spectrograph (HRS) instrument attached to the 9.2 m Hobby-Eberly-Telescope. A \ion{Na}{i} excess absorption of \mbox{0.0672 $\pm$ 0.0207 \%} was found, as well as a blue-shift in the planetary absorption signal within the transmission spectrum. The authors speculated that this was caused by high-speed winds flowing from the hot dayside to the cooler nightside of HD189733b. This result was confirmed by \citet{Wyttenbach2015} and \citet{CasasayasBarris2017} using the ESO 3.6 m telescope and the High Accuracy Radial velocity Planet Searcher (HARPS) spectrograph as well as \citet{Khalafinejad2017} using the Ultraviolet and Visual Echelle Spectrograph (UVES) instrument attached to the Very Large Telescope (VLT). Several investigations attempted to detect K on HD189733b. For instance, \citet{Jensen2011} used the Hobby-Eberly-telescope, but without success. A tentative 2.5-$\sigma$ detection of \ion{K}{i} in the atmosphere of HD189733b is shown by \citet{Pont2013} where the Advanced Camera for Surveys (ACS) instrument on-board of the Hubble Space Telescope (HST) was used for their observations. Recently, \citet{Keles2019} detected \ion{K}{i} with an excess level of \mbox{0.184 $\pm$ 0.022 \%} on HD189733b using the Large Binocular Telescope (LBT) and the Potsdam Echelle Polarimetric and Spectroscopic Instrument (PEPSI).

A particular emphasis is given to wind properties within the atmosphere of HD189733b, which are investigated in different high-resolution transit observations e.g. \citet{Louden2015}, \citet{Brogi2016}, \citet{Salz2018} or \citet{Seidel2020}. Winds on giant exoplanets can vary, e.g. being equatorial jets that originate from the lower parts of an atmosphere (around 1 bar) and stream into the direction of the planet's rotation, day-to-night-side winds which arise at higher altitudes (around 1 mbar) or atmospheric super-rotation \citep{Brogi2016}. These winds can vary from planet to planet, as different effects have an influence on their atmospheric properties e.g. a planetary magnetic field or synchronization between the atmosphere and planet interior \citep{Brogi2016}. In the case of day-to night-side winds, the line cores of resolved absorption lines show a blue-shift as the wind pushes the atmosphere into the observer's line of sight during transit. In the other cases, the absorptions arising from different positions on the observable annulus are velocity shifted regarding the projected distance to the rotation axis (largest shift on the greatest projected distance and lowest at the region where the atmosphere becomes opaque). On the trailing and leading limbs the shift is of opposite direction \citep{Louden2015}. The resulting observed final absorption line is the sum of each of these intrinsic shifted absorptions, which results in a broadened and shallower absorption line compared to an absorption line from a stagnant atmosphere. The aforementioned studies show a tentative consensus for the existence of an eastward equatorial jet in the atmosphere of HD189733b, which is consistent with modeling approaches (see \citet{Showman2013} and the references therein). 

In this work, we spectrally resolve the atmospheric \ion{K}{i} line on HD189733b from the observation presented by \citet{Keles2019} and compare the line shape to the resolved \ion{Na}{i}-D-lines presented by \citet{CasasayasBarris2017}. The alkali atoms Na and K are known to have very similar condensation curves as well as similar ionization potentials \citep{Lavvas2014}, leading to very similar shaped mixing ratio profiles \citep{lavvas2017}. As the observations suggest a strong broadening of the \ion{Na}{i}-D-lines, also recently shown by \citet{Seidel2020} and \citet{GebekOza2020}, we investigate the line widths by comparing them to synthetic line profiles.  We discuss the difference in the broadening of the alkali lines and their possible trigger, such as winds. The comparison between the alkali lines enables us to discuss either a scenario where the alkali lines probe the same atmospheric region which would hint on a super-solar planetary Na/K- ratio or a scenario where both features trace significantly different atmospheric regions with different temperature.

This paper is structured as follows. Section~\ref{sec:Data Analysis} shows the observational data and Section~\ref{sec:Models} the models used. The methods and results are presented in Section~\ref{sec:Results}, where we resolve the \ion{K}{i}-absorption and compare both alkali lines with synthetic transmission spectra. In Section~\ref{sec:Discussion}, we compare the alkali lines with each other and discuss the results. Section~\ref{sec:Summary} presents the summary and conclusion.

\section{Observational data}
\label{sec:Data Analysis}
\subsection{The \ion{K}{i} Observation}
A transit of HD189733b was observed at the LBT with PEPSI, which is a white--pupil fiber--fed spectrograph \citep[]{Strassmeier2015,Strassmeier2018,Strassmeier22018}. The observation covered the wavelength range \mbox{7340 - 9070 \AA} at a 4-pixel spectral resolution of 130\,000. During the observation, 15 out-of-transit and 9 in-transit spectra were observed at an exposure
time of 10 min. The telluric and blaze correction, as well as the image processing steps, are described in detail in \citet{Keles2019} (hereafter KEL19). The authors found a significant \ion{K}{i} excess absorption for bandwidths between \mbox{0.8 - 8.0 \AA} around the \ion{K}{i}-line at 7698.98 \AA{} in their observation. We use the processed and telluric corrected dataset from KEL19 and spline them to the wavelength grid with a spacing of 0.01 \AA{} to resolve the planetary \ion{K}{i}-line at \mbox{7698.98 \AA}. The second resonant \ion{K}{i}-line at \mbox{7664.92 \AA} is contaminated by strong telluric oxygen lines and is not investigated. 

\subsection{\ion{Na}{i} detection by \citet{CasasayasBarris2017}}
\label{sec:Detection by CSB}
\citet{CasasayasBarris2017} (hereafter CSB17) presented three transit observations of HD189733b acquired with the HARPS instrument \citep{Mayor2003} with a resolution of $\sim$115\,000 on the ESO 3.6 m telescope (La Silla Observatory). They comprised of 99 spectra (46 being in-transit and 53 being out-of-transit) at exposure times of 5 min and 10 min. The authors find a significant \ion{Na}{i} absorption resolved around both \ion{Na}{i}-D-lines at 5889.906 \AA{} and \mbox{5895.879 \AA}. Although the detection of the \ion{Na}{i}-D-lines using the same data set is made first by \citet{Wyttenbach2015}, we will compare the resolved \ion{K}{i}-line profile with the findings by CSB17, as the authors considered the effect of center-to limb-(CLV) variation \citep[see e.g.][]{Csesla2015,Yan2017} and the Rossiter-McLaughlin (RM) -effect \citep{Rossiter1924,McLaughlin1924} in their investigation. For this, we use the unbinned resolved \ion{Na}{i}-D-line profiles from CSB17 presented in their Figure 7 (top panel).

\section{Models}
\label{sec:Models}

\subsection{The residual RM-feature}
\label{sec:RM-effect}
Due to the stellar rotation, the stellar lines become shifted in wavelength according to their position on the stellar surface, whereby the line-shifts decrease towards the stellar center. During the transit, in case of HD189733, the planetary body blocks blue-shifted light during the first part of the transit and red-shifted light during the second part of the transit hence introduces a line deformation and shift to the observed mean stellar line. This effect introduces a residual feature after dividing the in-transit spectra by the master out-of-transit spectrum, which we denote here as the "residual RM-feature". 

The shape of the residual RM-feature in the transmission spectra depends strongly on the line-properties of the investigated stellar line, the star-planet geometrical properties and the stellar properties e.g. stellar rotation velocity and obliquity. This side effect introduced by the RM-effect is addressed in different investigations in the literature e.g. \citet{Louden2015}, \citet{Brogi2016}, \citet{Chen2020} and especially in the recent work by \citet{casasayasbarris2020}, showing the non-detection of \ion{Na}{i} in the atmosphere of HD209458b and the importance to consider the RM-effect.

We model the residual RM-effect using a grid model, where we map the stellar surface by a grid of 100 x 100 pixels containing limb angle dependent fluxes, thus considering also the CLV-effect (for further information on the stellar spectra we refer to section 3.3 in KEL19). To account for the stellar rotation, each spectrum is shifted according to the differential velocity on the stellar surface. During the transit, the planetary surface (mapped by 70 x 70 pixels) blocks different regions of the stellar surface, leading to the afore-mentioned effect. 

\subsection{The rotational broadening model}
\label{sec:line Broadening}
We model the rotational broadening of the synthetic transmission lines using a grid model, where we map the planetary surface by a grid of \mbox{100 x 100} pixels containing the synthetic lines. To account for the rotational broadening, each spectrum is shifted according to the differential velocity on the planetary surface. The planetary surface consists of the planetary body (at \mbox{1 $\times$ R\textsubscript{p}}) and the probed atmospheric ring, where the absorption lines originate. To compute the broadened absorption line profile, we sum over all pixels on the atmospheric ring. We define the border of the atmospheric ring at \mbox{1.285 $\times$ R\textsubscript{p}} for the \ion{Na}{i}-D2-line absorption, at \mbox{1.200 $\times$ R\textsubscript{p}} for the \ion{Na}{i}-D1-line absorption and at
\mbox{1.136 $\times$ R\textsubscript{p}} for the \ion{K}{i}-line absorption inferred according to the line contrasts of the non-broadened line profiles derived in this work (see Section~\ref{sec:Discussion}). Note, that these boundaries are first order approximations. However, the choice of the boundary has a small effect on the outcome, as the line broadening is determined mainly by the rotational velocity.
\newline
We validated our model by comparing it to the \footnote{\url{https://github.com/sczesla/PyAstronomy}} PyAstronomy tool \textit{pyasl.fastRotBroad}, which introduces rotational broadening according to \citep{Gray1994} (i.e. for a stellar surface). We introduced different broadening values to synthetic line profiles, which show no significant deviations from each other for both models.

\subsection{The Synthetic Transmission spectra}
We will compare the alkali lines with line profiles from synthetic transmission spectra. The theoretical transmission spectra are calculated for different atmospheric temperatures and volume mixing ratios. The temperature is varied between \mbox{2000 - 6000 K} in steps of \mbox{200 K}. The volume mixing ratio for potassium is varied in the range around \mbox{$\sim$ 10\textsuperscript{-9} -10\textsuperscript{-5}} and the sodium volume mixing ratio in the range of \mbox{$\sim$ 10\textsuperscript{-7} - 10\textsuperscript{-3}} around the solar composition values, which are $\sim$10\textsuperscript{-6} for Na and $\sim$10\textsuperscript{-7} for K \citep{Pino2018}. The atmosphere of HD189733b is well known for showing a scattering slope in the visible wavelength region that is far larger than predicted by molecular Rayleigh scattering alone \citep[e.g.][]{Pont2013}. A well-accepted explanation for this effect is the occurrence of small aerosol particles in the upper atmosphere that contribute to the scattering optical depth in this wavelength range. We follow the work of \citet{Pino2018} (see their equation (10)) and include this aerosol scattering via analytic approximations, which are based on the HST spectra of HD189733b. The transmission spectra are calculated with a reference pressure of 10 bar located at the measured white-light radius of HD189733b. Further details on the computation is provided in Appendix~\ref{sec:Synthetic Transmission spectra}.

All synthetic spectra were calculated with a spacing of 0.01\AA{} similar to the spacing of the observational data and convolved with the instrumental resolution using the PyAstronomy tool \textit{PyAstronomy.pyasl.instrBroadGaussFast}. As the synthetic transmission spectra mirror the wavelength dependent radius (R\textsubscript{p}) of HD189733b, we convert them via \mbox{1 -(R\textsubscript{p}/R\textsubscript{*})\textsuperscript{2}}, equivalent to the observational line profiles and normalize them with a second order polynomial fit to the continuum from 5000\AA{} - 10000\AA.
\begin{figure}
    \includegraphics[width=0.9\columnwidth]{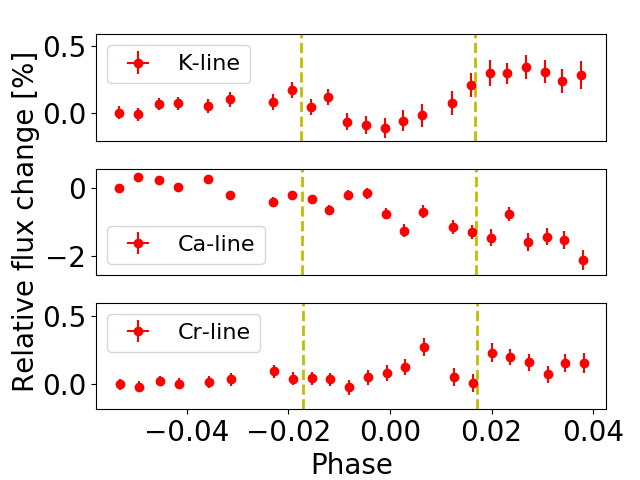}
    \caption{Excess absorption around a 0.5 \AA{} bandwidth centered around the \ion{K}{i}-line at 7699 \AA{} (first panel), the \ion{Ca}{ii}- line at 8662 \AA{} (second panel) and the control \ion{Cr}{i}-line at 7462 \AA{} (third panel).}
    \label{fig:Kel.png}
\end{figure}
\begin{figure*}
    \includegraphics[width=0.9\textwidth]{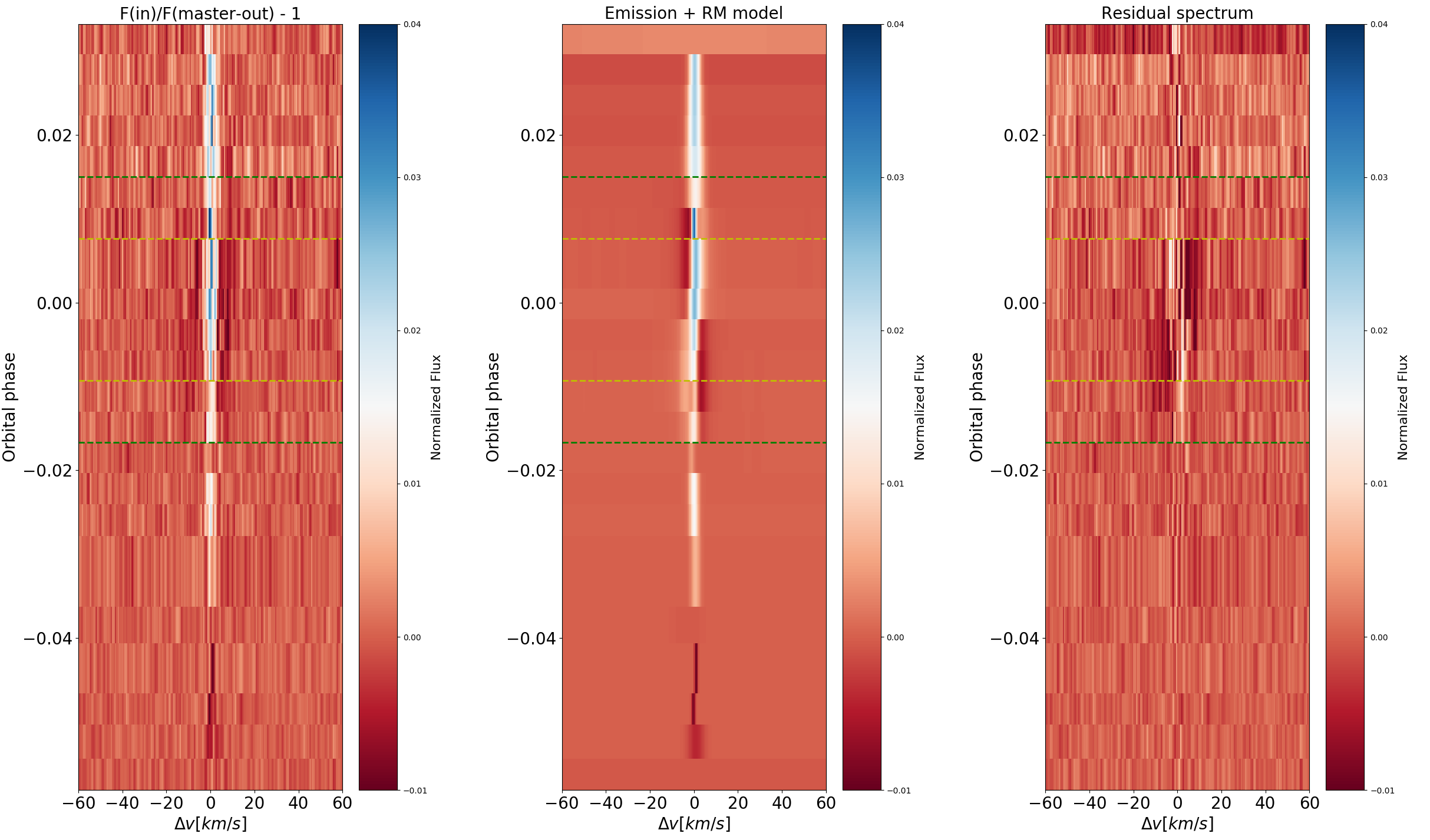}
    \caption{Transmission spectra around the \ion{K}{i}-line at 7698.98 \AA{} (orbital phase vs. velocity shift). Left panel: The observations divided by the Master out spectrum and subtracted by unity (red). Middle panel: The residual RM-model and the fit to the emission line induced by stellar activity. Right: The residual spectrum (left panel divided by the middle panel) showing the \ion{K}{i} absorption (dark region). Green dashed lines mark the 1st and 4rd contact and the yellow lines the 2nd and 3rd contact.}
    \label{fig:Residuals.png}
\end{figure*}

\section{Method and Results}
\label{sec:Results}
\subsection{Re-investigating the \ion{K}{i}-absorption on HD189733b}
\label{sec:Re-investigating the K-absorption on HD189733b}
We aim to spectrally resolve the planetary \ion{K}{i}-line and compare its line shape to the \ion{Na}{i}-D-lines presented in CSB17. Many processes can affect high-resolution transit spectra and need to be considered e.g. telluric lines (see e.g. \citet{Allart2017}), the center-to-limb variation (see e.g. \citet{Csesla2015},\citet{Yan2017}), the Rossiter-McLaughlin effect (see e.g. \citet{Cameron2010}, \citet{Cegla2016}, \citet{Dravins2018}, \citet{Bourrier2018}) or stellar activity (see e.g. \citet{CauleyKuck2018}). To separate the planetary \ion{K}{i}-absorption from the stellar line profile, one needs to derive the transmission spectrum by dividing the in-transit spectra (the observations acquired during the transit) by a master out-of-transit spectrum (the mean spectrum of the observations acquired before and after the transit). We derive the master out-of-transit spectrum using 14 out-of-transit spectra, neglecting ingress and egress observations. The planetary motion introduces a Doppler shift to the wavelength position of the planetary absorption lines. Due to this, each transmission spectrum is shifted back to the planetary rest frame and the transmission spectra co-added, demonstrating the planetary absorption. The left panel in Figure~\ref{fig:Residuals.png} shows the transmission spectra subtracted by unity (orbital phase vs. velocity shift). There is an emission-like feature increasing with planetary orbital phase, which is continuing also in the out-of-transit observations (visible beyond the green dashed lines which mark the 1st and 4rd contact). This is mentioned also by KEL19 and was related to stellar activity. 

In order to investigate spurious stellar activity during the transit, Figure~\ref{fig:Kel.png} shows the excess absorption around the \ion{K}{i}-line at \mbox{7699 \AA{}} (top), the \ion{Ca}{ii}-line at 8662 \AA{} (second panel) and the control \ion{Cr}{i}-line at 7462 \AA{} (third panel). The excess absorption derived here is the flux integrated around the line cores within a bandwidth of 0.5 \AA{} and normalized using a linear fit to the out-of-transit values. To show the relative flux change, the values are divided again by the value of the first observation. As mentioned by KEL19, we see a slope in the top two excess curves which can be well described by a linear fit. The \ion{Ca}{ii}-IR triplet lines are known to show variations due to stellar chromospheric activity. \citet{Klocova2017} investigated a flare during one transit of HD189733b determining a change in flux of the \ion{Ca}{ii}- line at 8498 \AA{} of around \mbox{2-3 \%} (see their Figure 6). Comparing this to Figure~\ref{fig:Kel.png}, we see a decrease in flux of around \mbox{2 \%} in the \ion{Ca}{ii}- line, showing a possible correlation with the activity. We show this correlation also for the weaker control \ion{Cr}{i}-line, which is less affected than the strong \ion{K}{i}-line. 

To account for the emission feature and the residual RM-feature (see Section~\ref{sec:RM-effect}) in the transmission spectra shown in the left panel of Figure~\ref{fig:Residuals.png}, we fit the emission line with a Gaussian profile using the Pyastronomy tool \textit{funcfit} and combine it with the residual RM-feature model (middle panel in Figure~\ref{fig:Residuals.png}). Due to the low rotation velocity of 3.3 km/s of the host star HD189733, the residual RM-feature does not move much and overlaps with the emission feature in the line core region. As the transit duration of around $\sim$109 min is quite short, there are only five transmission spectra available after neglecting the ingress and egress spectra, where the planetary absorption moves around $\pm$ 8 km/s corresponding to $\pm$ 0.21 \AA.

\begin{figure}
    \includegraphics[width=0.93\columnwidth]{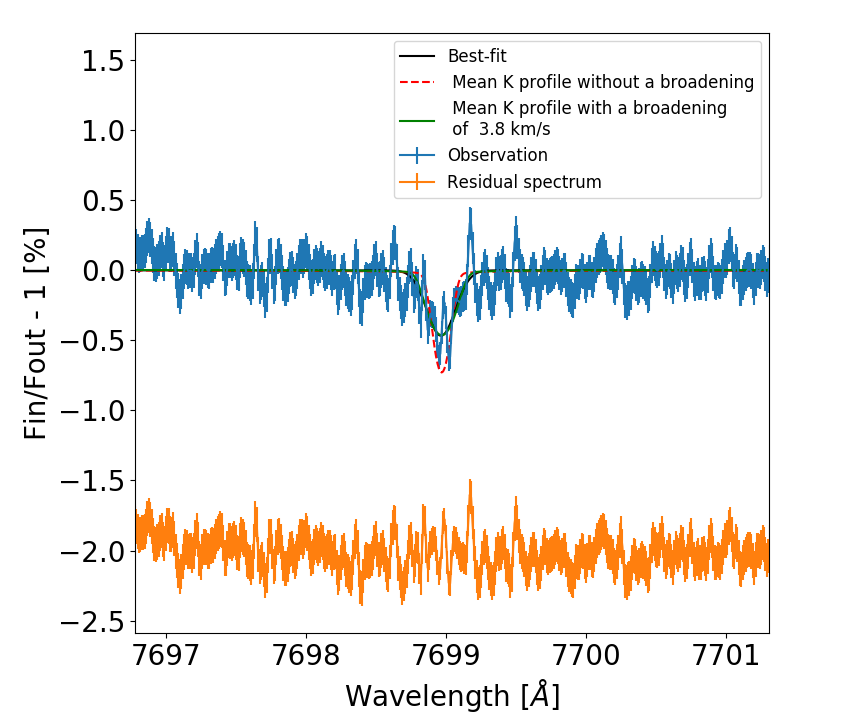}
    \caption{Transmission spectrum around the \ion{K}{i}-line at 7699 \AA{} in [\%] subtracted from unity. The black solid line shows the Gaussian fit. The dashed lines show the expected planetary absorption from the synthetic spectra without any broadening (red) and with the best-matching broadening solution of 3.8 km/s (green). The residual spectrum is shifted for clarity.}
    \label{fig:Kline.png}
\end{figure}

The transmission spectra are subsequently corrected for the resulting model by division (right panel in Figure~\ref{fig:Residuals.png}), showing the velocity shifted planetary absorption at different orbital phases. To derive the final planetary transmission spectrum, we shift the in-transit transmission spectra corrected by the model to the planetary rest frame and co-add them. The final transmission spectrum is shown in Figure~\ref{fig:Kline.png}. Applying a Gaussian fit (black solid line), we determine the FWHM = \mbox{0.24 $\pm$ 0.03 \AA} and the LC = \mbox{0.46 $\pm$ 0.04 \%}. The residual spectrum is shown with an offset (orange solid line), showing no correlated noise. We expect the line center at \mbox{7698.965 \AA{}} (as we removed the systemic velocity of -2.277 km/s). The line center is at \mbox{7698.966 $\pm$ 0.01 \AA} and shows no significant wavelength shift. 

To verify our result, we calculate the \ion{K}{i} excess absorption from the final transmission spectrum and compare it with the results of KEL19 for different integration bandwidths. For this, we integrate the final F\textsubscript{in}/F\textsubscript{out} spectrum around the line core and normalize it by the number of integrated points. The excess absorption agree within their error bars up to 4.0 \AA{} bandwidths and remain within 3-error bars for larger bandwidths. The deviations at larger bandwidths may arise due to the different techniques used to derive the excess absorption level \citep{casasayasbarris2020}.

\subsection{Comparison of the synthetic- and observational \ion{K}{i}-line profile}
\label{sec:Comparison K}
To investigate the \ion{K}{i}-line shape, we compare the resolved  \ion{K}{i}-line with the line profiles from the synthetic transmission spectra. For this purpose, we calculate the reduced $\chi$\textsuperscript{2}-value  \mbox{$\pm$0.24 \AA{}} around the \ion{K}{i}-line core corresponding to $\sim$2 $\times$ the FWHM (in total 48 spectrally resolved data points). Before the reduced $\chi$\textsuperscript{2}- value calculation, we introduce rotational broadening to the synthetic line profiles according to Section~\ref{sec:line Broadening} to account for line broadening. We increase this in steps of 0.1 km/s and calculate the reduced $\chi$\textsuperscript{2}-value repeatedly until we reach a reduced $\chi$\textsuperscript{2}-minimum between the observational and synthetic line profiles. We do this exercise for each synthetic temperature-abundance spectrum separately. 
\newline
Here, we emphasize that high resolution observations are insensitive to the atmospheric continuum and can only measure the contrast of an absorbing feature i.e. the difference between the line core and the continuum. This contrast does not change increasing (or decreasing) the abundance of a species, as the feature originates higher up (or lower down) in the atmosphere not changing its contrast \citep{Pino2018}. In this work, the continuum (and thus the contrast) is set by the haze layer. Therefore, increasing the temperature or abundance shifts the absorption line higher up in the atmosphere which arises more and more above the haze layer and mimics an increase in absorption for higher temperature and abundance.

\begin{figure}
    \includegraphics[width=\columnwidth]{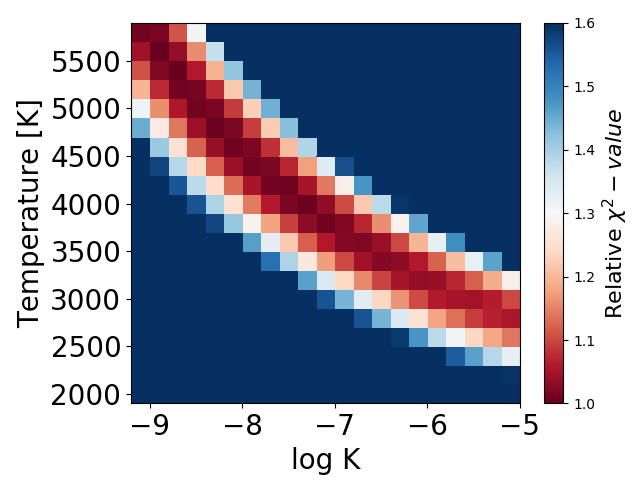}
    \caption{Shown is the abundance-temperature degeneracy for \ion{K}{i}-line. log K denotes the logarithmic volume mixing ratio of K. Reduced $\chi$\textsuperscript{2} values (color marked) relative to the best value compare the observational \ion{K}{i}-line with various synthetic line profiles. Each pixel corresponds to one comparison.}
    \label{fig:Comppot.png}
\end{figure}

The comparison map between the synthetic and observed \ion{K}{i} line profile is shown in Figure~\ref{fig:Comppot.png}. Each pixel corresponds to one synthetic transmission spectrum and observation comparison. The color bar shows the reduced $\chi$\textsuperscript{2}-minimum value relative to the best value. The characteristic shape of the color map highlights the degeneracy between the temperature and the K abundance. Increasing the temperature, the scale height increases as well and shifts the absorption to higher altitudes. Similarly, higher abundances lead to a stronger absorption signature above the haze layer which reaches higher altitudes. Thus, an absorption feature induced by an increased temperature can not be distinguished apriori by an absorption feature induced by higher abundance if the continuum is determined by a haze layer. The same scenario is valid for decreasing the temperature or abundance where the absorption is shifted to lower altitudes. Thus, increasing the temperature, synthetic spectra with lower abundances match the observations and show a lower reduced $\chi$\textsuperscript{2}- value (and vice versa). Note, that this effect is completely degenerate with the location of the haze layer and reference pressure level. Thus, the characteristic shape will shift linearly changing these parameters, denying the determination of the "real" absolute abundance - temperature values. 

However, from this grid of different temperature-abundance spectra, we are able to derive the mean synthetic \ion{K}{i}-line profile. For this, we determine the best matching line profile at each temperature grid on the characteristic shape (i.e. the red stripe) and calculate the mean line profile by averaging those. The mean synthetic line profile is shown in Figure~\ref{fig:Kline.png} with a green dashed line, while the mean synthetic \ion{K}{i}-line profile prior to the broadening process is presented in a red dashed line. A rotational broadening of $\sim$3.8 km/s is needed further to the accounted line broadening mechanisms such as instrumental resolution and atmospheric properties (for instance temperature and pressure broadening) to match the observational line profile (excluding the planetary rotation).
\begin{figure*}
    \includegraphics[width=0.9\textwidth]{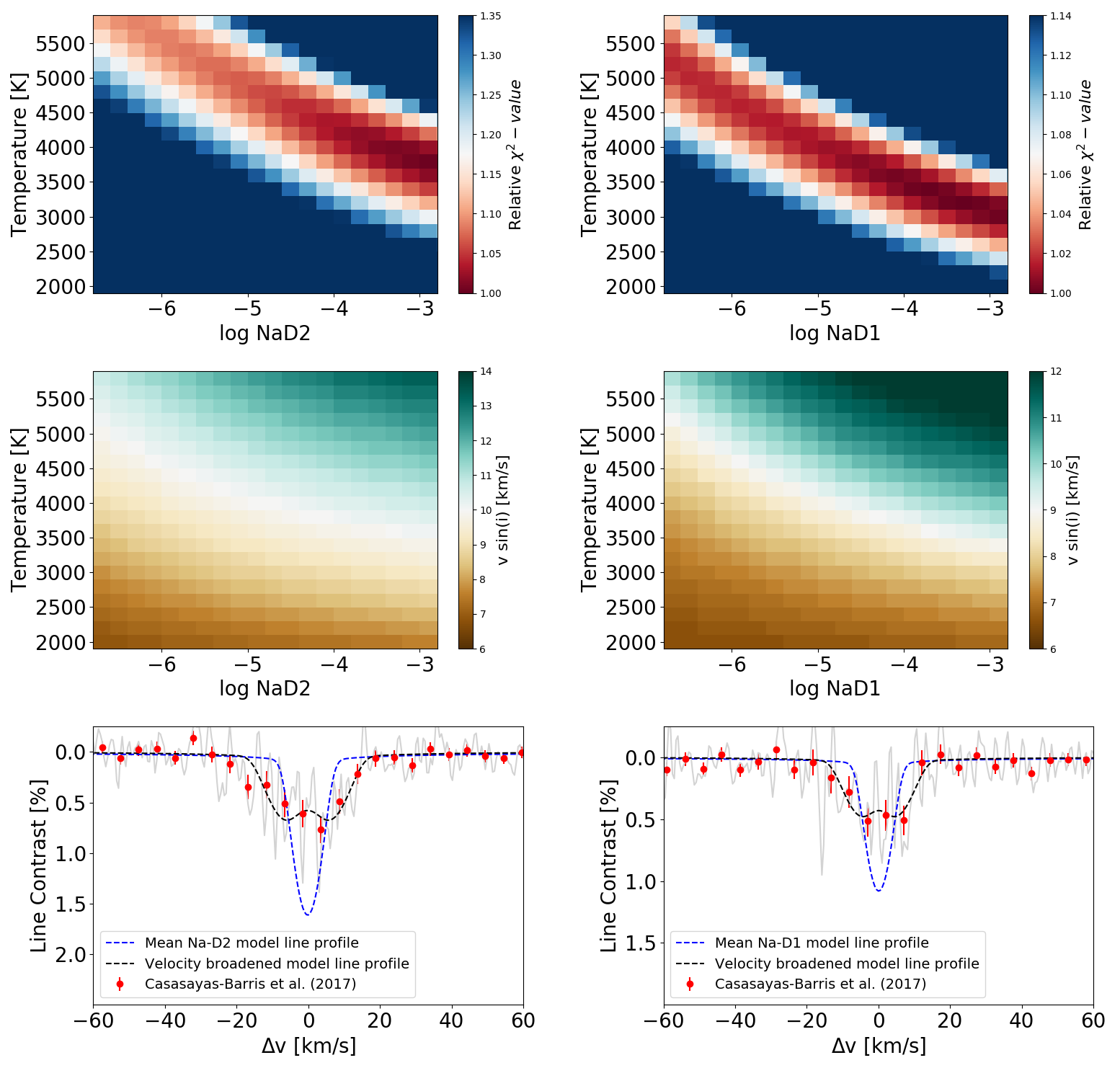}
    \caption{Top panel: Abundance-temperature degeneracy for \ion{Na}{i}-D-lines, where log NaD2 and log NaD1 are the logarithmic volume mixing ratios. Reduced $\chi$\textsuperscript{2} values relative to the best value (color marked) compare the observational \ion{Na}{i}-D-profiles with various synthetic spectra, whereby each pixel corresponds to one spectrum comparison. Middle panel: The introduced broadening to the synthetic spectra is shown. The color bar indicates the broadening value in km/s. Bottom panel: The line contrast over line center distance in units of velocity is shown for the\ion{Na}{i}-D2-line (left) and \ion{Na}{i}-D1-line (right). The synthetic mean line profile with (dashed black) and without (dashed blue) introduced broadening. Red dots show the resolved planetary \ion{Na}{i}-D-lines from \citet{CasasayasBarris2017} (binned by 10 pixels) and grey show their unbinned data.}
    \label{fig:Comparison.png}
\end{figure*}

\subsection{Comparison of the synthetic- and observational \ion{Na}{i}-D-lines}
\label{sec:Comparison of the synthetic with the observational NaD-lines}
We want to compare the resolved \ion{Na}{i}-D-lines in CSB17 (see Section~\ref{sec:Detection by CSB}) with the \ion{Na}{i}-D- absorption lines produced from the synthetic transmission spectra (in the same way as done in Section~\ref{sec:Comparison K} for the \ion{K}{i}-line). We compare the synthetic and observed absorption lines $\pm$0.62 \AA{} around the \ion{Na}{i}-D-line cores corresponding to \mbox{$\sim$2 $\times$ the FWHM} (in total 124 spectrally resolved data points for each \ion{Na}{i}-D-line).

Figure~\ref{fig:Comparison.png} shows the resulting comparison map for the \ion{Na}{i}-D-lines (top). Similar to the \ion{K}{i} investigation, one can see the degeneracy for the different temperature-abundance combinations due to the haze layer. The middle panel of Figure~\ref{fig:Comparison.png} shows the corresponding velocity broadening map, where the color bar indicates the introduced rotational broadening in km/s (the scales are adapted for best visibility). The maps show, that the synthetic lines of interest need to be broadened around $\sim$8-12 km/s to match the observation for both \ion{Na}{i}-D-lines. We derive the mean synthetic \ion{Na}{i}-D-line profiles in the same way as done for the mean synthetic \ion{K}{i}-line profile. The bottom panel of Figure~\ref{fig:Comparison.png} shows the resolved line profiles from CSB17 (unbinned = grey solid; binned by 10 pixels for clarity = red dots) and the derived mean synthetic \ion{Na}{i}-D-line profiles (dashed black line). The emission-peak in the line center arises due to the fact, that only broadened lines from the atmospheric ring are contributing to the final line profile, neglecting the blocked velocity shifted absorption profiles from the planetary body.
For small velocities, this peak vanishes, but becomes significant for larger ones. The mean synthetic broadened line profiles match well with the observations. We also show the mean synthetic line profiles before broadening (dashed blue).

\section{Discussion}
\label{sec:Discussion}
\begin{figure*}
    \includegraphics[width=\textwidth]{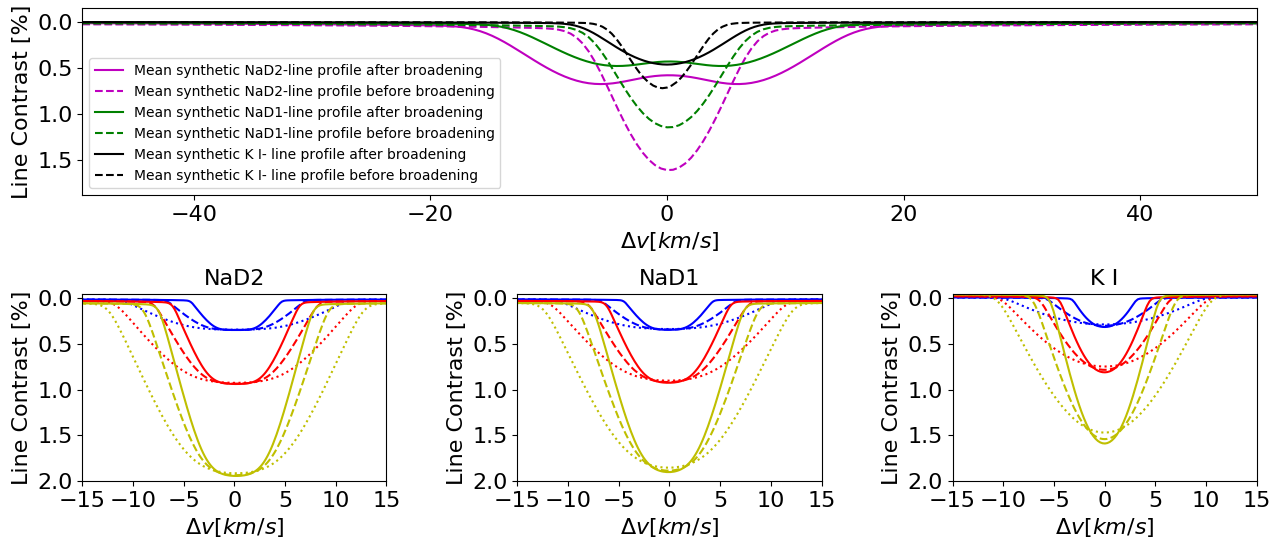}
    \caption{Top panel: Comparison of the mean synthetic \ion{Na}{i}-D-lines and the \ion{K}{i}-line from this work. Dashed lines show line profiles before and solid lines after introduced rotational broadening. Bottom panel: \ion{Na}{i}-D-lines and \ion{K}{i}-line modelled for isothermal temperature and solar abundance value at 2000K (blue), 4000K (red) and 6000K (yellow) for turbulence velocities of v\textsubscript{t} of 1 km/s (solid), 10 km/s (dashed) and 20 km/s (dotted). The x-axis shows the distance to the line center in km/s.}
    \label{fig:Turbulence.png}
\end{figure*}

We discuss the different possibilities, which could be the reason for the large difference in the line widths (around a factor of 2) for both alkali lines shown in the top panel of Figure~\ref{fig:Turbulence.png}, where we show the derived mean synthetic line profiles from this work. The comparison of the resolved \ion{Na}{i}-D-line profiles from CSB17 with the synthetically derived ones suggests that the \ion{Na}{i}-D-lines need to be broadened by velocities in the order of \mbox{$\sim$10 km/s}, regardless of the atmospheric temperature for the best matching synthetic line profiles, as shown in the middle panel of Figure~\ref{fig:Comparison.png}. Applying a Gaussian fit, the line profiles show a FWHM of \mbox{$\sim$22.9 km/s} for the \ion{Na}{i}-D2-line and \mbox{$\sim$18.8 km/s} for the \ion{Na}{i}-D1-line. Comparing the resolved \ion{K}{i}-line profile (which has an FWHM of \mbox{$\sim$9.4 km/s}) with the synthetically derived mean line profile, only a broadening of around \mbox{3.8 km/s} is needed, which is 2-3 times less compared to that we see for the \ion{Na}{i}-D-lines. The resolved \ion{Na}{i} and \ion{K}{i} lines were obtained with different spectrographs (HARPS and PEPSI). Both instruments are high resolution fiber-fed spectrographs which are pressure and temperature controlled and observed the alkali lines at similar resolution. The difference in line widths seems to be too large to be explained by potential systematic effects arising from the usage of different instruments. In both cases, one would expect a line broadening of around \mbox{2.7 km/s} due to the planetary rotation (further to the temperature and pressure broadening introduced into the synthetic line profiles). Taking into account that the alkali lines absorb at higher altitudes around 1-1.3 R\textsubscript{P}, this would increase the broadening around a few hundred m/s only. Furthermore, as the planet moves during the transit, the planetary absorption arises at different wavelength positions due to the Doppler-shift. At large exposure times, this introduce also a broadening to the observed line profiles regarding the orbital phase. For a 10 min exposure time, this is on the order of a few km/s (at the ingress and egress phase) and decrease down to a few hundred m/s at the mid-transit phase. The line broadening of the resolved \ion{K}{i}-line can be explained by the orbital motion and the fixed exposure time. 

Different atomic lines are found and resolved in other hot-and ultra-hot Jupiter type planets showing different line widths e.g. on WASP-49 \citep{Wyttenbach2017}, on Wasp-76 \citep{Seidel2019}, on Wasp-33b \citep{Yan2019}, on KELT-9b \citep{YanHenning2018, Yan2019, Hoeijmakers2019, Cauley2019,Turner2020}. To the best of our knowledge, so far only \citet{Chen2020} detected another resolved \ion{K}{i}-line for the exoplanet WASP-52b observed with the high resolution spectrograph ESPRESSO (Echelle SPectrograph for Rocky Exoplanets and Stable Spectroscopic Observations) at the VLT. The authors show the detection of the the \ion{Na}{i}-D-lines with a FWHM(\ion{Na}{i}-D2) = \mbox{21.6 $\pm$ 1.8 km/s} and FWHM(\ion{Na}{i}-D1) =  \mbox{11.5 $\pm$ 1.6 km/s} and the \ion{K}{i}-D1- line with a FWHM( \ion{K}{i}-D1) = \mbox{13.7 $\pm$ 3.3 km/s}. Even if the FWHM of the \ion{Na}{i}-D-lines show a strong deviation to each other, their mean FWHM is comparable with the FWHM of the \ion{K}{i}-D1-line, opposite to that we see comparing the \ion{K}{i}-line with the much broader \ion{Na}{i}-lines in our investigation. Large line widths do not have to be necessarily attributed to being broadened by e.g. winds, as also other factors determine the line broadening such as pressure- and temperature-broadening or the cross-section of the absorbing species. However, if the observed line broadening can not be explained by only accounting for these mechanisms, broadening by planetary winds can be considered as the trigger for the enhanced broadening. 

An explanation for the strong broadening of the \ion{Na}{i}-D-lines is shown by \citet{Seidel2020}. The authors show that super-rotational winds, day-to night-side winds or vertical wind patterns can introduce strong broadening to the \ion{Na}{i}-D-lines in the upper atmosphere of HD189733b. The authors propose a possible scenario, where a planetary magnetic field in the order of 50G propels Na\textsuperscript{+}-ions via the Lorenz force from a super-rotational jet in the lower atmosphere up to the upper atmosphere, which recombines there to neutral Na at the same speed. 

Another explanation for the large broadening may be turbulences in the planetary atmosphere. In contrast to the introduced rotational broadening, turbulences would lead to a line broadening by an intrinsic change of the Doppler-width which can be described by:
\begin{equation}
  \sigma_\mathrm{D} = \frac{\nu_0}{c} \sqrt{\frac{k_\mathrm{B} T}{m} + v_t^2} \ , 
\end{equation}
The bottom panel in Figure~\ref{fig:Turbulence.png} shows the effect of turbulent velocities v\textsubscript{t} on the \ion{Na}{i}-D-lines and the \ion{K}{i} line modelled for solar Na and K abundance value (at a isothermal temperature of 2000K, 4000K and 6000K for v\textsubscript{t} of \mbox{1 km/s}, \mbox{10 km/s} and \mbox{20 km/s}). Introducing rotational broadening, the corresponding equivalent width of the absorption lines remain conserved, thus deeper lines become broader but shallower. On the other hand, regarding turbulences, this is not the case and the equivalent widths are not conserved. High turbulent velocities flatten the line profiles in the line core region and introduce a characteristic shape, especially at lower temperatures. But also in the case of turbulences, one would expect similar broadening effects for both alkali lines if the turbulences are not variable. Note, that a mix of rotational broadening and turbulences may be possible. In this case, the observed line widths would be affected by the degeneracy between rotational broadening, temperature and turbulences, making the investigation even more complex.

In two recent works, \citet{Seidel2020} and \citet{GebekOza2020} investigate also the broadened \ion{Na}{i}-D-lines on HD189733b increasing the Doppler-width of the lines. Both authors derive velocities even larger than presented in this study for the \ion{Na}{i}-D-lines. However, a direct comparison of the wind velocities is not possible as the winds are introduced in a different way (Doppler width broadening vs. rotational broadening). Changing the Doppler-width (i.e. using Equation 1), the absorption profile changes as a result of the different velocity distribution of the absorbing Na- atoms, where v\textsubscript{t} can be also denoted as the microturbulence. On the other way, in case of rotational broadening, the Na atoms in each atmospheric column (in the line of sight) move as a unit and produce an absorption profile which is velocity shifted with respect to the distance to the rotational axis. The final absorption profile is then the sum of all velocity shifted line profiles on the atmospheric ring. However, the results are similar in a qualitative way, stating that the \ion{Na}{i}-D-lines may experience strong broadening on HD189733b.

However, we emphasize if the broadening of the lines arises from winds or turbulences in the order \mbox{$\sim$10 km/s} and higher, this would hint on wind velocities larger than the sound speed. But as the atmospheric region where the \ion{Na}{i} and \ion{K}{i} lines form is below the exobase and thus collisionally dominated, supersonic winds are not possible.  Therefore, the inferred velocities should be considered cautiously and rather qualitatively hinting on some broadening mechanisms in the upper atmosphere than quantitatively.

In general, both alkali features were expected to show very similar mixing ratio profiles and to probe similar pressure levels (see Figure 7 \& 12 in \citet{lavvas2017}) for HD189733b, thus one could expect similar line broadening. But comparing the absorption level (at similar bandwidths) for \ion{Na}{i} shown by \citet{Wyttenbach2015} and \ion{K}{i} shown by KEL19, \ion{Na}{i} probes $\sim$27 scale heights and so much higher altitudes compared to \ion{K}{i} which probes $\sim$13 scale heights, deviating a factor of $\sim$2 (assuming that the same temperature region is probed). A roughly similar result is inferred comparing the LC of the non-broadened alkali line profiles shown in the top panel of Figure~\ref{fig:Turbulence.png}. Hence, Na seems to probe higher altitudes compared to K. However, if Na probes higher temperature around T\textsubscript{Na} $\sim$ 2 $\times$ T\textsubscript{K} (and therefore higher altitude level), the noted difference in scale heights would be in agreement with the expectations. As the atomic mass of K is $\sim$1.7 times larger than the atomic mass of Na, the Na atoms may be lifted into the upper atmosphere by hydrodynamic motion more easily than K atoms, explaining the difference in altitude. However, advanced modelling effort would be needed to proof this which is out of the scope of this work.

If the \ion{Na}{i} absorption probes much higher altitudes compared to the \ion{K}{i} absorption, this could indicate that some mechanism is leading to a broadening to the absorption lines with decreasing pressure level on HD189733b. This is also suggested comparing the \ion{K}{i}-absorption to other high resolution investigations regarding wind properties. In case of day-to night-side winds, the largest blue-shifts are found for the lightest elements which probe the lowest pressure regimes such as He-I \citep{Salz2018} (\mbox{-3.5 $\pm$ 0.4 km/s} at the mid-transit) up to the evaporating regime probed by the Ly-$\alpha$ line \citep{Lecavelier2012}, where probably evaporation plays also a significant role on such a shift. We find no significant blue-shift of the resolved \ion{K}{i}-line, where we deduce a shift of \mbox{0.001 $\pm$0.01 \AA{}} \mbox{($\sim$ 0.04 $\pm$ 0.4 km/s)}. However, this can not exclude the presence of a day-to night-side wind on HD189733b. This value is similar within 2$\sigma$ to the findings by \citet{Louden2015}, who found a velocity shift of \mbox{$-1.9^{+0.7}_{-0.6}$ km/s} investigating the \ion{Na}{i}-D-lines. \citet{Brogi2016} used the CRIRES (CRyogenic high-resolution InfraRed Echelle Spectrograph) instrument at the VLT to investigate the different molecular absorption features around 2.3 $\mu$m during the transit of HD189733b. The authors also probe a lower day-to-night side wind of {$-1.7^{+1.1}_{-1.2}$ km/s}, which is also consistent within 2$\sigma$ with our finding. In case of rotational broadening, \citet{Brogi2016} state a rotational velocity of \mbox{v\textsubscript{rot} = $3.4^{+1.3}_{-2.1}$ km/s}, which would be in agreement within 1$\sigma$ with the broadening value needed for the \ion{K}{i}-line of $\sim$3.8 km/s. The results of \citet{Brogi2016} have been confirmed by \citet{Flowers2019} analyzing the same dataset applying 3D GCM (general circulation model) simulations, also being in agreement with our broadening values in sense of wind speed (see their Table 1) and blueshift-values. As the molecular lines originate at lower altitudes compared to the alkali-lines, this appears to be consistent with the fact that some mechanism is producing significant line broadening at low pressure levels. Comparing the widths of the \ion{Na}{i}-D-lines with the H$\alpha$-line \citep{Cauley2015}, both features probe similar altitudes and show similar large widths \citep{Huang2017}, strengthening this picture.

Recently, \citet{GebekOza2020} compared the observational \ion{Na}{i}-D-lines shown by \citet{Wyttenbach2015} for HD189733b with synthetic transmission spectra computed for different scenarios such as a hydrostatic atmosphere and three evaporative scenarios being an escaping atmosphere, an outgassed cloud sourced by an exomoon and a torus representing circumplanetary material, showing a strong evidence that these lines probe optically thin regions in the atmosphere. Comparing the K-absorption by \citet{Keles2019} with their scenarios, the best match is given by a hydrostatic scenario probing an optically thick region. The authors note, that their optically thin scenarios are very sensitive to the planetary Na/K ratio, which can vary if this differs from the stellar one. We aim to probe the planetary Na/K ratio and compare this to the stellar Na/K abundance ratio, which we derive in Appendix~\ref{sec:The stellar Na and K abundance ratio}. The inferred stellar Na/K abundance ratio is log(Na/K) = $\sim$1.2 similar to the solar value \citep{Asplund2009}. Although the exoplanet could have undergone evolutionary processes changing its Na/K ratio, we expect a similar Na/K ratio for HD189733b which seems to be an appropriate assumption in the first order \citep{Lavvas2014}. Note, that due to the extinction by Earth's atmosphere, high-resolution transit observations are typically not suitable to determine absolute abundances of atmospheric constituents. The continuum information usually gets lost due to re-normalization of the spectra and becomes degenerate with the reference pressure level and transit radius \citep[]{BennekeSeager2012,Hengkitzmann2017,Welbanks2020}, whereby under certain circumstances the degeneracies can be broken (see e.g. \citet{BennekeSeager2012}, \citet{Fisher2018}, \citet{Brogi2019} or \citet{Welbanks2020}). However, the degeneracies cancel out (for the model assumptions used in this work) deriving the ratio of the abundances, such as the Na/K ratio, unless the temperature at which the lines form is different. We roughly estimate the Na/K ratio by comparing the upper panel of Figure~\ref{fig:Comppot.png} and Figure~\ref{fig:Comparison.png} assuming that both features arise at the same atmospheric temperature and use the abundance value for the best matching model (i.e. the model with the lowest reduced $\chi$\textsuperscript{2} value). For the temperature range of 3400K - 5000K, we infer a planetary $\overline{\ion{Na}{i}}/\ion{K}{i}$ ratio which is $\sim$ 20 - 400 times higher than the stellar Na/K abundance ratio. Note, that if Na originates at higher temperature than K, the planetary Na/K abundance ratio will decrease to a lower level, down to the solar Na/K abundance value. We emphasize here that the roughly estimated Na/K values should be considered only as a hint that the alkali lines may probe different atmospheric temperature regions, as the inferred Na/K ratio has large uncertainties and relies on different assumptions (which we present in Appendix~\ref{sec:The planetary Na/K abundance ratio}). Furthermore, such an increased abundance ratio seems to be unlikely to be present either in the protoplanetary disk or the atmosphere of the planet taking into account primordial elemental abundances and alkali chemistry as discussed for the hot Jupiter HD209458b in \citet{Lavvas2014}, which also orbits a solar metallicity star. Furthermore, a strong variation in the alkali abundances due to depletion by e.g. by ionization processes as well as condensation processes seems to be unreasonable. Condensation for K and Na would be expected below an atmospheric temperature of 1000 K for both species in a very similar way \citep{lavvas2017}, where both alkalis can condense into various different solids and liquids. Likely candidates (see e.g. \citep{Marley2013}) are sodium sulfide (Na2S) and potassium chloride (KCl). The former condenses around 900 K in a solar metallicity atmosphere at 0.1 bar, while the latter requires slightly lower temperatures (about 720 K). For a extended presentation of alkali chemistry we refer here to \citet{Lavvas2014}. In the case of photo-ionization, K ($\sim$4.34 eV) has only slightly lower ionization potential than Na ($\sim$5.14 eV) \citep{Fortney2003,Barmann2007}, making also this process unlikely to be the reason for such a difference.

\section{Summary \& Conclusion}
\label{sec:Summary}
We compared previously observed high resolution \ion{Na}{i} \citep{CasasayasBarris2017} and \ion{K}{i} \citep{Keles2019} absorption in the atmosphere of HD189733b with synthetic transmission spectra modeled for a variety of temperature and abundance values. The comparison of the \ion{Na}{i}-D-lines shows that the observed \ion{Na}{i}-D-line widths are much larger than the modeled ones. The \ion{Na}{i}-D-lines have to be broadened by velocities in the order of \mbox{$\sim$10 km/s} to match the observations if only rotational broadening is taken into account. To compare the \ion{K}{i}-line with the synthetic transmission spectra, we used the previous high-resolution observation by KEL19 and resolved the \ion{K}{i}-absorption from the data. The \ion{K}{i}-line profile shows a broadening comparable with the modeled synthetic line profiles, which is significantly less in comparison to the \ion{Na}{i}-D-lines. Comparing to different investigations, there is a hint that the line widths show stronger broadening with increasing altitude: Starting from molecular absorption signature from the lower atmosphere showing weak or even no significant broadening effects \citep{Brogi2016}, up to the slightly larger broadened \ion{K}{i}-line (this work) and even to higher altitudes where the \ion{Na}{i}-D-lines \citep{Wyttenbach2015,Louden2015,Barnes2016,CasasayasBarris2017,BorsaZannoni2018,Seidel2020} and the H$\alpha$-line \citep{Cauley2015} show very large broadening. The same picture is drawn comparing the wavelength shifts introduced by a probably day-to night-side wind, which shows stronger blue-shifts for the lines emerging at lower pressure levels. This hints that the main \ion{K}{i}-absorption may arise from lower altitudes than the \ion{Na}{i}-absorption. Estimating the planetary Na/K ratio and comparing this to the derived stellar Na/K ratio, HD189733b would possess a very high super-solar atmospheric Na/K ratio if Na and K trace the same atmospheric temperature for the model assumptions used in this work. However, the derived Na/K ratio depends on several assumptions and is such enhanced, that this scenario seems to be very unlikely. In case that K traces lower altitudes at a cooler temperature compared to Na, the planetary Na/K ratio can coincide with the stellar Na/K ratio being in agreement with other studies such as \citep{Welbanks2019}, which seems to be a more likely scenario.

Another high resolution and high S/N transit observation, which covers the wavelength range of different atomic and molecular species, would be needed to compare the broadening mechanisms to avoid artifacts attributed to the stellar activity, different instrumentation or even variable weather conditions on the exoplanet. 

\section*{Acknowledgements}
We sincerely thank N. Casasayas Barris for sharing the high resolution data including the resolved \ion{Na}{i}-D-lines.
\newline
The research leading to these results has received funding from the European Research Council (ERC) under the European Union's Horizon 2020 research and innovation programme (grant agreement No. 679633; Exo- Atmos).
\newline
KGS thanks the German Federal Ministry (BMBF) for the year-long support for the construction of PEPSI through their Verbundforschung grants 05AL2BA1/3 and 05A08BAC as well as the State of Brandenburg for the continuing support of LBT and PEPSI (see https://pepsi.aip.de/). LBT Corporation partners are the University of Arizona on behalf of the Arizona university system; Istituto Nazionale di Astrofisica, Italy; LBT Beteiligungsgesellschaft, Germany, representing the Max-Planck Society, the Leibniz-Institute for Astrophysics Potsdam (AIP), and Heidelberg University; the Ohio State University; and the University of Notre Dame, University of Minnesota and University of Virginia.
\newline
XA is grateful for the financial support from the Potsdam Graduate School (PoGS) in form of a doctoral scholarship.
\newline
JVS is supported by funding from the European Research Council (ERC) under the European Union's Horizon 2020 research and innovation program (project {\sc Four Aces}; grant agreement No. 724427).

\section*{Data Availability}
The data underlying this article will be shared on reasonable request to the corresponding author.



\bibliographystyle{mnras}
\bibliography{kelesbib} 



\appendix
\section{Synthetic Transmission spectra}
\label{sec:Synthetic Transmission spectra} 
We compute theoretical transmission spectra by calculating the effective tangent height and the wavelength-dependent, apparent planetary radius with the \texttt{Helios-o} model as described in \citet{Bower2019A&A...631A.103B} or \citet{Gaidos2017MNRAS.468.3418G}. The isothermal atmosphere is divided equidistantly in log(P) into 200 layers from 10 bar to 10$^{-10}$ bar. The spectra are computed at a constant resolution of 0.01 cm$^{-1}$ in the wavenumber range between 10,000 cm$^{-1}$ and 20,000 cm$^{-1}$.

The opacity sources that are included in the calculations are the collision-induced absorption of H$_2$-H$_2$ and H$_2$-He pairs, the D$_1$ and D$_2$ resonance lines of \ion{K}{i} and \ion{Na}{i}, and molecular Rayleigh scattering of H$_2$ and He. The line strengths $S$ for the \ion{Na}{i} and \ion{K}{i} resonance lines are obtained via
\begin{equation}
  S = \frac{g_2 A_{21}}{8 \pi \nu_0^2 Q} \exp\left(-\frac{E_1}{k_\mathrm{B} T} \right) \left[ 1 -\exp\left(-\frac{hc\nu_0}{k_\mathrm{B} T}\right)\right] \ ,
\end{equation}

where $g_2$ is the statistical weight of the upper level, $A_{21}$ the Einstein A-coefficient, $\nu_0$ the wavenumber of the transition, $E_1$ the energy of the lower level, $Q$ the partition function, and $T$ the temperature. The values for $g_2$ and $A_{21}$ are taken from \citet{Draine2011piim.book.....D}, while $E_1$ is zero because the resonance lines are ground-level transitions. The line profile is modeled by a Voigt profile, consisting of a Doppler core and Lorentzian line wings, where the standard deviation of the Doppler profile $\sigma_\mathrm{D}$ is given by
\begin{equation}
  \sigma_\mathrm{D} = \frac{\nu_0}{c} \sqrt{\frac{k_\mathrm{B} T}{m}} \ , 
\end{equation}

with the molecular mass of the species $m$.
The \ion{Na}{i} and \ion{K}{i} resonance lines are well known for possessing a far-wing line profile that is non-Lorentzian \citep[e.g.][]{Burrows2003ApJ...583..985B, Allard2016A&A...589A..21A}. We approximate this behavior by using the analytic fits to the \citet{Burrows2003ApJ...583..985B} line profiles according to \citet{Baudino2015A&A...582A..83B}. We note, however, that because of the strong aerosol scattering slope, the far wings of these lines are practically invisible and only the thermally-broadened Doppler cores are located above the spectrum's continuum level.

\section{Cautions on the estimated planetary Na/K abundance ratio}
\label{sec:The planetary Na/K abundance ratio}
The estimated planetary Na/K abundance ratio in this work is based on different assumptions, which will be discussed here. The temperature distribution at the terminator may not be constant as expected from the isothermal profiles, thus it can affect the abundance ratio, as the line cores form at a much higher temperature than the line wings \citep{Huang2017}. We estimate the Na/K ratio from the Figure~\ref{fig:Comppot.png} and Figure~\ref{fig:Comparison.png}, where we introduced rotational broadening. If the line broadening is a result of turbulences, the equivalent width of the absorption lines is not conserved, thus the line widths will be larger without an increase in line depth, which will lead to overestimation of the Na/K abundance ratio. Furthermore, the synthetic line profiles depend on the functional form of the continuum which is determined by the aerosol layer. Changing the functional form of the continuum will also affect the results presented here. For instance, \citet{Welbanks2019} derived the atmospheric abundances of Na and K for several planets and compared them to those of their host stars. In this study, the authors investigated also the planet HD189733b using HST data acquired with the STIS instrument as presented by \citet{Sing2016}. The authors infer similar stellar and planetary Na and K abundances based on their retrieval analysis of these data. However, inspecting panel 15 on their Figure 2, the retrieval results are based only on 2 data points around the \ion{K}{i}-line, resulting probably in a debatable conclusion. However, as the derived planetary Na/K abundance ratio is much larger than what would be expected, we only state the estimated super-solar Na/K ratio on HD189733b as a hint that Na and K may originate at different atmospheric regimes, as the aforementioned effects have a strong impact on the result, similar to non-LTE effects, which could also play a role. Note that high resolution observations are not always described well by simple theoretical models (such as the one employed here) and mismatches between observational and modelled line profiles can be found e.g. in \citet{Hoeijmakers2018} and \citet{Hoeijmakers2019}. While in those studies, the Fe lines match extremely well with the theoretical predictions, the Fe$^+$ measurements, on the other hand, deviate by quite a large degree. It is obvious, that the theoretical models lack some of the physics to describe the spectral lines or the chemical abundances in the very high atmosphere properly. 

\section{The stellar Na and K abundance ratio}
\label{sec:The stellar Na and K abundance ratio} 
\begin{figure*}
\includegraphics[width=\columnwidth]{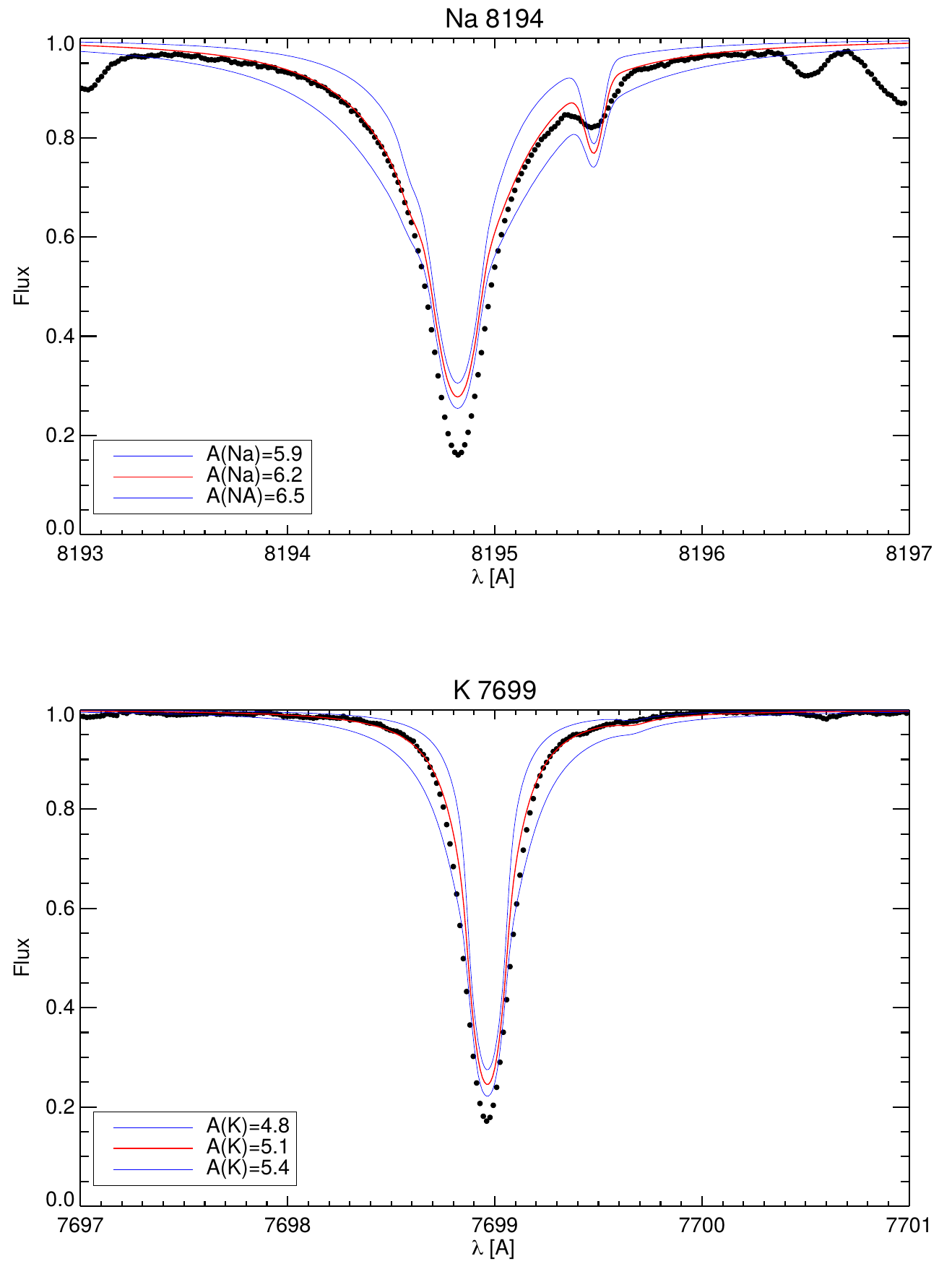}
    \caption{Comparison of Na-line at 8195 \AA{} (top) and the 7699 \AA{} \ion{K}{i}-line (bottom) from PEPSI spectra with the spectrum synthesis model based on a standard 1D model atmosphere. "A" denotes the logarithmic abundance value.}
    \label{fig:Na_K_abundance.png}
\end{figure*}

The metallicity of HD\,189733 is known to be roughly solar, according to previously published investigations, and so is its Na abundance, e.g. \mbox{[Na/H]=-0.04 $\pm$ 0.06}, \mbox{[Fe/H]=-0.10 $\pm$ 0.03} \citep{Montes2018}. However, even if the Na/K abundance ratio of HD189733 is expected to be solar \citep{Welbanks2020}, we could not find similar abundance information about K. We decided therefore to derive an estimate of the stellar K abundance based on the available PEPSI spectrum. For this purpose, we selected the clean \ion{K}{i}-line at 7699 \AA{} and compared its profile with synthetic spectra computed for a set of different K abundances, $A$(K)\footnote{$A$(X)$=\log(N$(X)$/N$(H))+12}, between $4.5$ and $5.5$, centered on the solar abundance of $A$(K)=$5.03\pm 0.09$ \citep{Asplund2009}. The comparison is shown in the bottom panel of Figure~\ref{fig:Na_K_abundance.png}. 

The spectrum synthesis is based on a standard 1D model atmosphere taken from the MARCS grid \citep{Gustaffson2008} with parameters $T_{\rm eff}=5000$\,K, $\log g=4.5$, [Fe/H]=0.0, and $\xi_{\rm micro}=1.0$\,km/s, which represents HD\,189733 reasonably well \citep[cf. stellar parameters compiled by][]{Southword2010}. This model atmosphere was then used by the line formation code Turbospectrum\footnote{\url{ttps://github.com/bertrandplez/Turbospectrum2019}} \citep{Plez2012} that generates synthetic line profiles for a given set of atomic parameters characterizing the spectral line in question. The atomic line data were taken from the VALD3 database\footnote{\url{http://vald.astro.uu.se/}} \citep[][and updates]{Ryabchikova2015}, including updated pressure broadening constants. 

Comparing observed and synthetic line profiles, we notice that the line cores of the synthetic spectra are too shallow. This presumably indicates that the assumption of LTE (Local Thermodynamic Equilibrium) is not a valid approximation in the core of this rather strong \ion{K}{i}-line (line depth $d> 0.8$). However, we may assume that the wings of the line, where $d<0.2$, form in LTE conditions and can be used to estimate the K abundance from a comparison with the synthetic LTE spectra. The best match of the wings is found for $A$(K)=$5.1\pm 0.1$. This result is fully consistent with the expectation that HD\,189733 has a solar K abundance. As a sanity check, we have performed the same exercise with the Na-line at 8194 \AA{}. The comparison is shown in the top panel of Figure~\ref{fig:Na_K_abundance.png}. Also, in this case, the observed line core is much deeper than the model results, for the same reason responsible for the mismatch in the \ion{K}{i}-line discussed above. Fitting the line wings, we deduce $A$(Na)=$6.2\pm 0.1$, which is fully consistent with the solar photospheric abundance of $A$(Na)=$6.24\pm 0.04$ \citep{Asplund2009}. We conclude that the stellar Na/K abundance ratio is indistinguishable from solar, $N$(Na)/$N$(K) $\approx 16$.


\bsp    
\label{lastpage}
\end{document}